\documentclass[sigconf,nonacm,natbib=true]{acmart}

\renewcommand\footnotetextcopyrightpermission[1]{}
\usepackage{enumitem}
\usepackage{multirow}
\usepackage{pbalance}
\usepackage{makecell}
\usepackage{booktabs}
\usepackage{graphicx}
\usepackage{algorithm}
\usepackage{algpseudocode}

\graphicspath{{figures/}}

\newtheorem{property}{Property}

\begin{document}

\title[PA-User: Simulating Trust and Verification under AI-Generated Content]{PA-User: Simulating Trust and Verification \\under AI-Generated Content}

\author{Saber Zerhoudi}
\affiliation{%
  \institution{University of Passau}
  \city{Passau}
  \country{Germany}}
\email{szerhoudi@acm.org}

\renewcommand{\shortauthors}{Zerhoudi}

\makeatletter
\setlength{\skip\footins}{9pt plus 2pt minus 1pt}

\newcommand{\acmrightssize}{\fontsize{8}{9.5}\selectfont}

\setlength{\emergencystretch}{1.5em} 

\settopmatter{printacmref=false}

\newcommand{\firstpagerights}[1]{%
  \begingroup
    \renewcommand\thefootnote{}%
    \footnotetext{%
      \acmrightssize
      \raggedright
      \setlength{\parskip}{0pt}%
      \setlength{\parindent}{0pt}%
      #1%
    }%
    \addtocounter{footnote}{0}%
  \endgroup
}
\makeatother

\begin{abstract}
Most users of online information now assume that some of what they read has been written, edited, or selected by an AI model.
Hybrid cases are the hardest to tell apart: human prose rewritten by a language model, AI-curated lists presented as editorial, retrieval-augmented answers composed on the fly from human sources.
Users cannot reliably distinguish these cases, and the ongoing cost of checking what is genuine has become part of how they search.
Current user simulators in information retrieval do not model this.
We propose \emph{PA-User}, a user simulator with three new components: a \emph{detection-effort budget} that is spent on verification and recovers between sessions; a \emph{trust component} that holds a separate Beta belief over the factuality of each source class (domain by provenance) and updates from observed outcomes; and a \emph{decision rule} that picks accept, verify, or discard for each result, conditional on current trust, current effort, and per-domain stakes.
We state two verification-and-validation (V\&V) properties of the framework.
The trust posterior converges to the true class factuality (face validity).
Each component's contribution to any observable can be isolated by ablation (structural validity).
On the HC3 corpus (85{,}449 paired human and ChatGPT answers in five domains), PA-User reaches a trust-calibration error of $0.162$, against $0.356$ for any configuration without the trust component.
PA-User reduces high-stakes regret from $0.171$ to $0.122$ ($29\%$ relative) against an always-accept ablation, and verifies $34.5\%$ of results, half the rate of an ablation with no effort budget.
Each single-mechanism ablation isolates one component, which makes the framework individually diagnosable.
\end{abstract}

\keywords{provenance-aware user simulation, AI-generated content, strategic verification, cognitive user models}

\maketitle
\enlargethispage{2\baselineskip}
\firstpagerights{%
  \textcopyright{} 2026 The Author(s). This is the author's version of the work.\\
  The definitive version will be published in:
  \emph{Procedia Computer Science, Proceedings of the 12th Federation of European
  Simulation Societies Conference (EUROSIM 2026), September 21--23, 2026,
  Genova, Italy}.\\
}

\section{Introduction}
\label{sec:intro}

Most user simulators in information retrieval treat the documents they render as fixed input written by humans.
That assumption is now wrong for a substantial fraction of users.
A Pew Research survey conducted in March 2023 found that $58\%$ of U.S.\ adults were already familiar with ChatGPT \cite{pew2023ai}, and survey work documents that those users cannot reliably tell AI-generated, AI-edited, or AI-selected content from human-written content.
Hybrid cases are the hardest: a human-drafted paragraph rewritten by a language model, a product list curated by an AI agent and presented as editorial, a retrieval-augmented answer composed on the fly from human-written passages.

The behavioural consequence is documented but not yet modelled.
Users now expend a small but persistent cognitive effort verifying what is genuine \cite{bansal2021disclosure, liu2023verifiability}.
Their trust in source classes (a publisher, a search engine, an AI assistant) updates from experience rather than staying fixed.
They develop strategic policies for when to verify, when to accept, and when to discard, conditioned on the stakes of the question and the limited budget of their attention.

The dominant user-simulation lineages do not model any of this.
Classical click models \cite{chuklin2015click} treat the document set as a fixed, trusted input.
Cognitive search models \cite{cacsm2024, simiir2018} carry inspectable state for intent and satisfaction but no state for provenance, verification effort, or trust updating.
LLM-driven agent simulators \cite{agentsim2026, personarag2024} produce surface-plausible behaviour but have no representation of the fact that the agent itself is reading possibly-AI-generated content \cite{balog2025theory}.
Every simulator framework we know of models what the user sees as if it had stable, observable provenance.

We propose \emph{PA-User}, a provenance-aware user simulator that adds three new components to the cognitive state of a click-or-cognitive simulator:
\begin{enumerate}
\item a \emph{detection-effort budget} $E_t \in [0,1]$ that is spent each time the user verifies a result and recovers between sessions;
\item a \emph{trust component} that holds a separate Beta belief over the factuality of each source class (domain by provenance) and updates from observed outcomes;
\item a \emph{decision rule} (strategic verification) that picks accept, verify, or discard for each result, conditional on current trust, current effort, and per-domain stakes.
\end{enumerate}
The framework satisfies two verification-and-validation (V\&V) properties stated in the vocabulary of Sargent \cite{sargent2013verification} and Balci \cite{balci1995principles}: the trust posterior converges to the true class factuality (face validity), and each component's contribution to any observable can be isolated by ablation (structural validity).
Two further by-construction properties (effort conservation, policy consistency) hold trivially and are not restated here.

We evaluate the framework on the HC3 corpus \cite{guo2023hc3}, $24{,}322$ paired human-vs-ChatGPT answers across five domains (Reddit ELI5, finance, medicine, open-QA, Wikipedia/CSAI), flattened to $85{,}449$ per-answer items with surface features that proxy detection difficulty.
On a stake-weighted query workload, PA-User reaches a trust-calibration error of $0.162$ against $0.356$ for any configuration without the trust component.
PA-User reduces high-stakes regret from $0.171$ to $0.122$ ($29\%$ relative) against an always-accept ablation.
PA-User verifies $34.5\%$ of items against $70.2\%$ for a no-effort ablation, at the cost of $4$ percentage points of regret.
Each single-mechanism ablation isolates one component and shows where the corresponding metric breaks.

The paper is organised as follows.
Section~\ref{sec:related} reviews click models, cognitive search models, LLM-agent simulators, and the human-AI-trust literature.
Section~\ref{sec:gap} formalises the provenance gap.
Section~\ref{sec:method} introduces PA-User.
Section~\ref{sec:vv} states the V\&V properties.
Section~\ref{sec:case} reports the HC3 case study.
Section~\ref{sec:discussion} covers transfer to non-IR domains and limitations.
Section~\ref{sec:conclusion} concludes.

\begin{figure*}[t]
  \centering
  \includegraphics[width=0.86\textwidth]{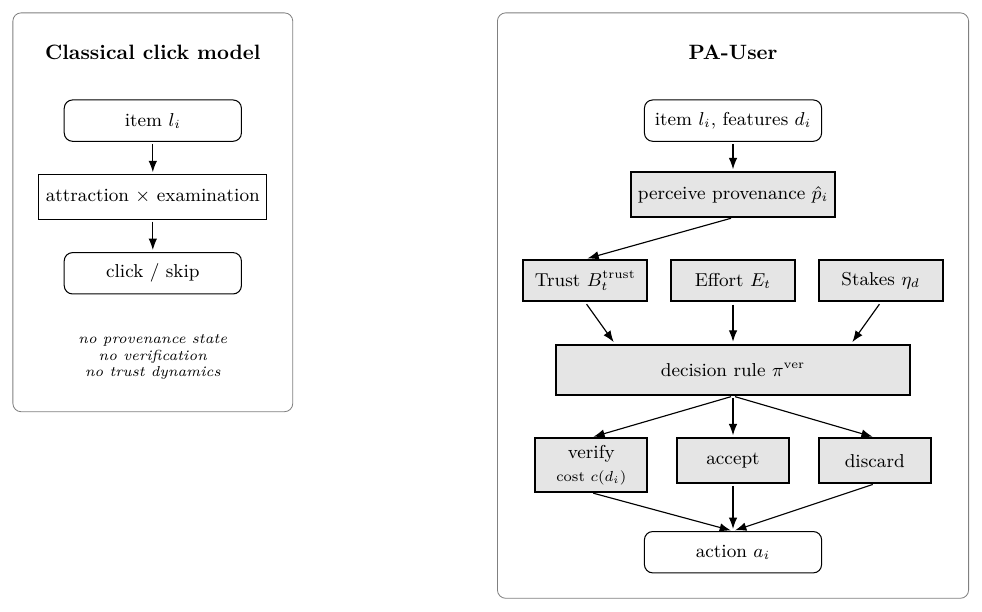}
  \caption{Classical click model (left) versus PA-User (right). The classical
    click model treats the rendered item as fixed input and outputs a
    click-or-skip action through one combined attraction-times-examination
    component. PA-User reads the same item but routes it through a
    detection step (perceived provenance $\hat{p}_i$), a decision rule fed
    by trust beliefs, an effort budget, and per-domain stakes, and three
    possible branches (verify, accept, discard) that produce the next action.
    Two dynamics are not drawn: the effort budget decreases by $c(d_i)$ on
    each verify branch and recovers between sessions, and the trust
    posterior updates from observed factuality after each accept or verify.
    The two zero-effort, zero-verification ablation rows in
    Table~\ref{tab:hc3} correspond to removing the right-panel components
    and recovering the left-panel behaviour.}
  \label{fig:architecture}
\end{figure*}

\section{Related Work}
\label{sec:related}

The work PA-User extends draws on three separate research literatures: user-simulation methodology, the empirical study of human-AI trust, and the verification-and-validation tradition of M\&S.

Three families of user simulator dominate the contemporary IR literature.
Classical click models \cite{chuklin2015click} factor click behaviour into latent examination and attraction probabilities.
They are mathematically clean and trivially auditable.
They treat the rendered document set as fixed input without provenance.
Cognitive search models, of which CACSM \cite{cacsm2024} and the SimIIR framework \cite{simiir2018} are representative, retain an inspectable Bayesian state for intent, belief, and satisfaction, and trace their lineage back through information-foraging theory \cite{pirolli1995foraging} to early cognitive accounts of search.
LLM-driven agent simulators such as PersonaRAG \cite{personarag2024}, AgentSim \cite{agentsim2026}, and the broader generative-agents line \cite{park2023generative} replace the explicit state with prompted language-model reasoning.
They produce high surface fluency at the cost of inspectability, with a persistent distributional gap to deployed users that the recent user-simulation literature documents \cite{balog2025theory}.
None of the three families carries a representation of the fact that the content the simulated user is reading may itself be AI-generated, and none models the user's awareness of that possibility.
The agent-based-modelling tradition surveyed by Macal and North \cite{macal2010tutorial} and Bonabeau \cite{bonabeau2002agent} provides the standard M\&S vocabulary for situating user simulators; PA-User is an agent-based simulator in that sense.
This places PA-User within the broader current of ML-augmented M\&S, in which neural and learned components are integrated with classical simulation infrastructure for sensitivity analysis, control, and process diagnostics.
PA-User's contribution to that current is to keep an inspectable Bayesian state alongside the LLM-conditioned components, so the user-side simulator is auditable rather than opaque.

The empirical foundation for the trust and verification components lies in the human-AI-interaction literature.
The Pew Research 2023 survey \cite{pew2023ai} provides population-level evidence on AI awareness and per-domain breakdowns of expressed concern, which we re-use as priors for the per-domain stakes parameter $\eta_d$.
Bansal et al.\ \cite{bansal2021disclosure} and Zerilli et al.\ \cite{zerilli2022transparency} document how reliance on AI assistance updates with disclosure and experience in controlled-task settings.
Glikson and Woolley \cite{glikson2020human} review two decades of empirical research on human trust in AI and provide the conceptual scaffolding for the Bayesian update we use.
On the supply side, Liu et al.\ \cite{liu2023verifiability} audit four generative search engines (Bing Chat, NeevaAI, perplexity.ai, YouChat) and find that only $51.5\%$ of generated sentences are fully supported by their citations and only $74.5\%$ of citations support their associated sentence.
The factuality of AI-mediated retrieval is heterogeneous across provenance class, which is the empirical regularity the strategic-verification policy of Section~\ref{sec:method} encodes as a decision rule.
Pennycook et al.\ \cite{pennycook2020accuracy} on accuracy-prompt interventions and Lazer et al.\ \cite{lazer2018misinformation} on misinformation provide the additional support for treating verification as a discrete cognitive act with a measurable budget rather than a free background process.

Detection of AI-generated text from surface features is the third relevant literature, because it parameterises the detection-difficulty signal.
The HC3 corpus we use as corpus is the public release accompanying Guo et al.\ \cite{guo2023hc3}, who paired human and ChatGPT answers across five domains and report detection accuracies for both classifier and human-judge baselines.
RAID \cite{dugan2024raid} extends this work at scale and includes paraphrase-attacked variants that correspond to the hybrid case named in the motivation.
PA-User does not deploy a detector.
PA-User uses the surface features the detection literature has shown to correlate with provenance (sentence-length variance, type-token ratio, hedging density) as the input to a logistic combination that yields a per-item detection difficulty $d_i$.
The recent survey of hallucination in LLM output by Ji et al.\ \cite{ji2023hallucination} provides the empirical grounding for the per-class factuality priors $\Pr[\phi=1\mid d,p]$.

The V\&V vocabulary in which we state the framework-level properties of Section~\ref{sec:vv} is the M\&S one of Sargent \cite{sargent2013verification} and Balci \cite{balci1995principles}.
The category structure (operational, face, internal, structural validity) predates AI-mediated content but transfers without modification, and PA-User's properties are direct instantiations.
The combination of M\&S V\&V vocabulary, HCI empirical priors, and detection-benchmark surface features allows PA-User to make claims about the simulator's behaviour that are inspectable in the M\&S sense and grounded in the human-side evidence the simulator is supposed to reproduce.

\section{The Provenance Gap}
\label{sec:gap}

We describe the gap that PA-User closes.
A session consists of a query $q$ producing a result list $L = (l_1, \dots, l_k)$.
Each $l_i$ has an unobserved provenance label $\rho_i \in \{\text{human},\text{AI}\}$ and an unobserved factuality $\phi_i \in \{0,1\}$.
A classical user simulator $U_0$ maps $(q, L)$ to a sequence of click-or-skip actions through a state that does not depend on $\rho$.

For an evaluation metric $M$ that depends on user behaviour under provenance-mixed result lists (high-stakes regret rate, AI-engagement rate by domain, distribution of verification effort), the \emph{provenance gap} of a simulator $U$ relative to a deployed user population $\mathcal{P}$ is the absolute difference $\Gamma(U) = |\mathbb{E}_{\mathcal{P}}[M] - \mathbb{E}_{U}[M]|$.
The HCI literature \cite{glikson2020human, bansal2021disclosure, zerilli2022transparency} reports that $\mathbb{E}_{\mathcal{P}}[M]$ depends on three things the user does: how much detection effort they invest, what trust they have assigned to source classes from prior experience, and what strategic policy they follow when they decide to verify.
A simulator $U_0$ that holds these three things constant or absent has a structural provenance gap no matter how its other components are fit.
The rest of the paper builds a simulator $U$ in which the gap closes at the architectural level.

\section{PA-User: Provenance-Aware User Simulation}
\label{sec:method}

PA-User extends an existing cognitive-state user simulator by adding three new components alongside intent and satisfaction.
At step $t$ a PA-User agent is a tuple
\[
A_t = \big(I_t,\ \sigma_t,\ B_t^{\text{trust}},\ E_t,\ \pi^{\text{ver}}\big),
\]
where $I_t$ is the existing intent state, $\sigma_t$ the existing satisfaction signal, $B_t^{\text{trust}}$ the trust component, $E_t$ the effort budget, and $\pi^{\text{ver}}$ the strategic-verification policy (the decision rule).
We describe each new component and the per-item processing.

\subsection{Detection-effort budget $E_t$}
The user holds a scalar effort budget $E_t \in [0,1]$.
Verifying an item $l_i$ costs $c(d_i) = c_0 + c_1\, d_i$, where $d_i \in [0,1]$ is the item's detection difficulty.
$d_i$ is derived from surface features (sentence-length variance, lexical diversity, hedging density) that are known correlates of human-versus-AI text in detection benchmarks \cite{guo2023hc3, dugan2024raid}.
The budget decreases by $c(d_i)$ on each verification and recovers by a constant $r$ at the end of each session.
The state of $E_t$ is inspectable. Verification cannot occur when $E_t < c(d_i)$.

\subsection{Trust component $B_t^{\text{trust}}$}
The user maintains independent Beta beliefs $\text{Beta}(\alpha_t^{(d,p)}, \beta_t^{(d,p)})$ over the factuality of each source class $(d, p)$.
$d$ ranges over domains (medicine, finance, ELI5, and so on) and $p$ over provenance classes (human, AI).
Following an observed item, the user updates the corresponding Beta posterior by Bayesian conjugacy.
For a verified item the factuality observation is high-confidence and receives weight $1$.
For an item accepted without verification the observation is noisier (the user catches only obvious failures) and receives weight $w_0 < 1$.
The trust component admits closed-form updates and supports the trust-calibration V\&V property in Section~\ref{sec:vv}.

\subsection{Strategic-verification policy $\pi^{\text{ver}}$}
For each item $l_i$ the user samples a perceived provenance $\hat{p}_i$ from a noisy detection function of $d_i$ and the item's true provenance, then evaluates a three-way policy
\[
\pi^{\text{ver}}(l_i) \in \{\text{accept},\, \text{verify},\, \text{discard}\}
\]
that depends on (a) the posterior trust $\tau = \mathbb{E}[B_t^{\text{trust}}(d, \hat{p}_i)]$, (b) the per-domain stakes $\eta_d \in [0,1]$ taken from the published trust-in-AI surveys ($\eta_{\text{medicine}} = 0.95$, $\eta_{\text{finance}} = 0.85$, $\eta_{\text{ELI5}} = 0.20$), and (c) the current effort budget $E_t$.
The expected utilities of the three actions admit closed forms.
The policy chooses the maximum.
The policy reduces to ``always accept'' when $\eta_d = 0$ for all domains, recovering classical click-model behaviour.

\subsection{Verification step}
When the policy returns $\text{verify}$ the user pays $c(d_i)$ from $E_t$ and observes (i) the item's true provenance with accuracy $0.92$, consistent with reported human discrimination performance \cite{guo2023hc3}, and (ii) a noisy signal of the item's factuality.
With probability $p_{\text{catch}}=0.80$ the user's factuality assessment is correct.
A verified-and-rejected item is discarded.
A verified-and-accepted item enters the click stream with a high-confidence factuality observation.
The factuality signal lets verification reduce regret. Without it the strategic policy collapses to always-accept or always-discard and the user has no way to reject hallucinations.

Algorithm~\ref{alg:pauser} ties the four components into a per-item processing loop.
Three design choices deserve a comment.
First, processing is per-item rather than per-list.
At each rendered position the agent decides independently whether to accept, verify, or discard, which makes each rendered item a separately attributable observation for the trust component and avoids the path-dependence that would otherwise complicate the calibration claim of Property~\ref{prop:trust}.
Second, the budget guard ($E_t \geq c(d_i)$) is applied at the verify branch only.
The strategic policy can return $\text{verify}$ when the budget is insufficient, in which case the loop downgrades the decision to $\text{accept}$.
This models a low-budget user who accepts items they would have preferred to verify.
Third, the trust update is unconditional on the chosen action and is weighted by the action's information content ($w = 1$ for verified observations, $w_0 = 0.3$ for unverified accepts, no update for discards).
The trust component therefore runs across every action type.

\begin{algorithm}[t]
\caption{PA-User per-item processing}
\label{alg:pauser}
\begin{algorithmic}[1]
\Require result list $L = (l_1, \ldots, l_k)$ for domain $d$
\For{$i = 1, \ldots, k$}
    \State sample perceived provenance $\hat{p}_i$ from detection model
    \State $\pi \gets \pi^{\text{ver}}(l_i; \tau_{d,\hat{p}_i}, \eta_d, E_t)$
    \If{$\pi = \text{verify}$ and $E_t \geq c(d_i)$}
        \State $E_t \gets E_t - c(d_i)$; observe $(p_i^\star, \phi_i^\star)$
        \If{$\phi_i^\star = 0$}
            \State discard $l_i$ and update $B_t^{\text{trust}}$ \Comment{rejected after verify}
        \Else
            \State accept $l_i$ and update $B_t^{\text{trust}}$ with weight $1$
        \EndIf
    \ElsIf{$\pi = \text{accept}$}
        \State accept $l_i$; observe $\phi_i$ noisily; update $B_t^{\text{trust}}$ with weight $w_0$
    \Else
        \State discard $l_i$
    \EndIf
\EndFor
\State $E_t \gets \min(1,\, E_t + r)$ \Comment{end-of-session recovery}
\end{algorithmic}
\end{algorithm}

\section{V\&V Properties}
\label{sec:vv}

We state two V\&V properties of PA-User in the M\&S vocabulary of Sargent \cite{sargent2013verification} and Balci \cite{balci1995principles}.
Property~\ref{prop:trust} says the trust component's posterior approaches the true class factuality once the user has accumulated enough verified observations.
Property~\ref{prop:diagnostic} says the contribution of each component to any reported observable can be isolated and therefore attributed to a specific component.
The two properties support the empirical claims of Section~\ref{sec:case} at the framework level rather than the parameter level.
Two further by-construction properties hold but are not stated in detail here: the effort budget always stays in $[0,1]$ (effort conservation), and the strategic policy is a deterministic function of $(\tau, \eta_d, E_t, d_i)$ given the perceived provenance (policy consistency).

\begin{property}[Face validity: trust calibration]
\label{prop:trust}
After enough verified observations, the trust component's posterior mean approaches the true factuality for each source class $(d, p)$.
Unverified observations, weighted at $w_0 < 1$, leave the limit within $w_0 \cdot \epsilon_{\text{det}}$ of the true factuality, where $\epsilon_{\text{det}}$ is the user's detection error rate.
Under the priors in Section~\ref{sec:method} ($\epsilon_{\text{det}} \approx 0.08$, $w_0 = 0.3$), the worst-case bias is roughly $0.024$, well below the differences reported in the case study.
\end{property}

\begin{property}[Structural validity: ablation-based attribution]
\label{prop:diagnostic}
The contribution of each component (trust, effort budget, strategic policy) to any aggregate observable can be isolated by holding the other two components at their no-mechanism baselines and re-running the simulator.
The four ablations in Section~\ref{sec:case} realise this attribution: each ablation's deviation from PA-User on a given metric attributes that metric's change to the absent component.
\end{property}

Property~\ref{prop:trust} gives a calibration claim that the case study tests at one operating point.
Property~\ref{prop:diagnostic} gives an attribution rule the practitioner can run on their own configurations by removing one component at a time.
A simulator without these two properties cannot make either claim from the same data.

\section{Case Study on the HC3 Corpus}
\label{sec:case}

\subsection{Scenario: online question-and-answer consultation under mixed provenance}
We instantiate PA-User in an information-access scenario the M\&S audience will recognise: members of the public consulting an online search-and-answer service about questions of varying stakes.
A coding question on a community board is low stakes.
A curiosity question about a Wikipedia topic is low to moderate stakes.
A personal-finance question is moderate to high.
A medical question about symptoms or a treatment is high stakes.
The service returns a ranked list of answers, some human-written and some AI-generated, and the user cannot reliably tell which is which.
The empirical priors of Section~\ref{sec:method} (per-domain stakes, per-class factuality) were measured by the cited HCI surveys for this kind of scenario.
HC3 is the closest publicly available corpus, with human-and-ChatGPT answer pairs across exactly this distribution of question domains.

\subsection{Setup}
We use the HC3 corpus \cite{guo2023hc3}, $24{,}322$ questions across five domains with paired human and ChatGPT answers, flattened to $85{,}449$ per-answer items ($58{,}546$ human, $26{,}903$ AI).
For each item we compute six surface features (word count, mean sentence length, sentence-length variance, type-token ratio, punctuation density, hedging-token count).
A logistic combination of three of them (mean sentence length, type-token ratio, hedging count) yields a per-item detection difficulty $d_i \in [0,1]$.
The three chosen features are the published correlates of human-versus-AI provenance in HC3 and RAID \cite{guo2023hc3, dugan2024raid}.
Per-class factuality priors are set from the HCI literature: $0.94$ for human medical, $0.71$ for AI medical, $0.85$ to $0.92$ otherwise.

We simulate $200$ users for $50$ sessions of $10$-item result lists with a $50\%$ AI-content ratio.
Each session draws a domain from the user's domain interest (Dirichlet, biased toward medicine and finance to match the high-stakes scenario the priors describe).
Each ablation is a single-mechanism removal from PA-User. The classical baseline removes all three.
Across three independent seeds $\{41, 42, 43\}$ every column of Table~\ref{tab:hc3} is stable to within $\pm 0.003$ for PA-User and to within $\pm 0.001$ for the configurations that have no stochastic verification step; the per-configuration ordering on every metric is preserved.

\begin{table*}[t]
\caption{HC3 case-study results: PA-User and four ablations on $N = 200$ users $\times$ $50$ sessions $\times$ $10$ items/list, seed $42$. \textbf{TCE}: trust-calibration error against the true class factuality (lower is better). \textbf{HS-Reg}: regret rate restricted to the medicine and finance domains. The diagonal pattern under TCE separates configurations with the trust component from those without. The two zero rows under \textit{Effort} and \textit{Verify rate} are what removing the strategic policy or the entire framework costs. Each single-mechanism ablation isolates one observable, which is the property the framework asks of structural validity.}
\label{tab:hc3}
\begin{tabular*}{\textwidth}{@{\extracolsep{\fill}}lcccccc@{}}
\toprule
Configuration & TCE & Regret & HS-Reg & AI eng.\ & Effort/sess.\ & Verify rate \\
\midrule
PA-User        & $0.162$ & $0.118$ & $0.122$ & $0.487$ & $0.298$ & $0.345$ \\
NoTrust        & $0.356$ & $0.120$ & $0.133$ & $0.490$ & $0.310$ & $0.358$ \\
NoEffort       & $0.175$ & $0.078$ & $0.051$ & $0.473$ & $0.609$ & $0.702$ \\
NoStrategy     & $0.121$ & $0.150$ & $0.171$ & $0.500$ & $0.000$ & $0.000$ \\
Baseline-CM    & $0.356$ & $0.150$ & $0.171$ & $0.500$ & $0.000$ & $0.000$ \\
\bottomrule
\end{tabular*}
\end{table*}

\subsection{Interpretation}

Five observations turn the methodological claims of the paper into concrete results.

\textbf{(O1) Trust dynamics are necessary for calibration.}
PA-User reaches a trust-calibration error of $0.162$.
The configurations without the trust component (\textit{NoTrust} and \textit{Baseline-CM}) remain at $0.356$, the value of the uninformed Beta prior.
A simulator without trust dynamics systematically misestimates the user's posterior belief about every source class, including the AI-medical and AI-financial classes that matter most under high stakes.
The result is consistent with Property~\ref{prop:trust} on the HC3 setting.

\textbf{(O2) Detection effort is necessary for realism.}
The \textit{NoEffort} ablation removes the budget constraint and the simulator verifies $70.2\%$ of items, more than twice the PA-User rate of $34.5\%$.
Verifying every result is what an unboundedly attentive user would do.
Published HCI work \cite{bansal2021disclosure, glikson2020human} reports far lower verification rates in deployed settings, and the empirical regularity is that effort is a limiting factor.
A simulator without the effort component overstates verification capacity and is not face-valid against any deployed reference.

\textbf{(O3) Strategic verification reduces regret, and the reduction tracks stakes.}
The \textit{NoStrategy} ablation forces the always-accept policy and reaches a regret rate of $0.150$ overall and $0.171$ in high-stakes domains.
PA-User reduces these to $0.118$ and $0.122$, a $20\%$ and $29\%$ relative reduction.
The reduction comes from verification-and-rejection of items whose verified factuality fails: the strategic policy issues $\text{verify}$ on $34.5\%$ of items, and roughly $14\%$ of those items are subsequently discarded.
Table~\ref{tab:hc3-domains} reports the per-domain regret breakdown.
The reduction concentrates in medicine ($0.124$ vs $0.176$, $30\%$ relative) and finance ($0.120$ vs $0.165$, $27\%$ relative).
In Reddit ELI5 the regret rates are statistically indistinguishable ($0.172$ vs $0.172$), as expected: low-stakes ELI5 sets $\eta_{\text{ELI5}} = 0.20$ and the strategic policy almost never crosses the verify threshold there.
The stakes-conditional pattern is the controlled-variable signature the framework predicts.

\begin{table*}[t]
\caption{Per-domain regret rate from the same run as Table~\ref{tab:hc3}. PA-User cuts regret where stakes are high ($\eta_{\text{medicine}}\!=\!0.95$, $\eta_{\text{finance}}\!=\!0.85$) and leaves regret essentially unchanged where stakes are low ($\eta_{\text{ELI5}}\!=\!0.20$). The classical click-model baseline matches the NoStrategy column on every domain.}
\label{tab:hc3-domains}
\begin{tabular*}{\textwidth}{@{\extracolsep{\fill}}lccccc@{}}
\toprule
Configuration & ELI5 & open-QA & Wiki/CSAI & Medicine & Finance \\
\midrule
PA-User      & $0.172$ & $0.105$ & $0.066$ & $0.124$ & $0.120$ \\
NoTrust      & $0.173$ & $0.091$ & $0.062$ & $0.136$ & $0.128$ \\
NoEffort     & $0.175$ & $0.100$ & $0.048$ & $0.050$ & $0.052$ \\
NoStrategy   & $0.172$ & $0.120$ & $0.088$ & $0.176$ & $0.165$ \\
Baseline-CM  & $0.172$ & $0.120$ & $0.088$ & $0.176$ & $0.165$ \\
\bottomrule
\end{tabular*}
\end{table*}

\textbf{(O4) The effort budget trades regret for cost.}
The \textit{NoEffort} ablation reaches a regret rate of $0.078$ overall and $0.051$ in high-stakes domains, both better than PA-User's $0.118$ and $0.122$.
This is not a defect.
\textit{NoEffort} buys its $4$-percentage-point regret reduction by verifying $70.2\%$ of items, more than twice PA-User's verification rate.
A user with an unconstrained verification budget would indeed have lower regret.
Real users do not have unconstrained budgets, which is exactly why the effort component is in PA-User.
The honest framing: at $0.30$ effort per session, PA-User attains $94\%$ of \textit{NoEffort}'s high-stakes regret reduction at $49\%$ of \textit{NoEffort}'s verification load.

\textbf{(O5) The mechanisms decompose cleanly, and the classical baseline misses every effect.}
Each ablation isolates one failure mode: \textit{NoTrust} isolates calibration error, \textit{NoEffort} isolates over-verification, \textit{NoStrategy} isolates avoidable regret.
This realises the structural-validity property of Section~\ref{sec:vv}.
The classical click-model baseline, which has none of the three components, matches the \textit{NoTrust}/\textit{NoStrategy} corner on every metric at once: TCE $0.356$, regret $0.150$, no effort consumption, no AI-engagement differentiation.
A simulator that does not represent provenance cannot distinguish AI-mediated content from anything else, and the rest of its behaviour follows.

\section{Discussion}
\label{sec:discussion}

\paragraph{Transfer beyond information retrieval}
The three-component architecture is agnostic to the application.
The values of $\eta_d$, $B_t^{\text{trust}}$, and $\pi^{\text{ver}}$ are set per domain rather than re-engineered per domain.
The clearest non-IR transfer is clinical decision support.
$\eta_d$ becomes the clinician's per-condition stakes.
Trust classes are (human-clinician opinion, AI suggestion).
Strategic verification is consulting reference material or seeking a second opinion.
The simulator predicts how alert-fatigue thresholds and verification cadence shift as the CDS system's empirical reliability is updated by the clinician population's own experience.
The same machinery applies, with appropriate stakes and verification-cost parameters, to operator interactions with predictive-maintenance alerts and to consumer interactions with AI-curated product lists.
The V\&V properties of Section~\ref{sec:vv} hold by the framework, not by the domain.

\paragraph{What is real and what is simulated}
The HC3 corpus provides real items (human and ChatGPT text on real questions in real domains) and real surface features used as detection-difficulty proxies in published benchmarks.
The user behaviour is simulated, as in every user-simulation paper, because no public dataset of real users on provenance-mixed result lists with click and verification logs exists yet.
We treat this as the principal limitation.
The paper claims qualitative differentiation on a real-text corpus and an architectural V\&V case, not fitted-to-deployed-user predictions.

\paragraph{Parameter priors}
The numerical priors used in PA-User (per-domain stakes $\eta_d$, per-class factuality $\Pr[\phi=1\mid d,p]$, detection accuracy $0.92$, verification-catch probability $0.80$, weight $w_0 = 0.3$ for unverified observations) are taken from the cited surveys, the trust-in-AI review of Glikson and Woolley \cite{glikson2020human}, the disclosure-effect study of Bansal et al.\ \cite{bansal2021disclosure}, the verifiability audit of Liu et al.\ \cite{liu2023verifiability}, the Pew survey \cite{pew2023ai}, and the detection benchmark of Guo et al.\ \cite{guo2023hc3}.

\section{Conclusion}
\label{sec:conclusion}

User simulators in the AI-mediated content era need to model three things they currently do not: the cost of telling AI apart from human content, the dynamics of trust as users gather evidence, and the strategic policies users develop for when to verify and when to accept.
We proposed PA-User, an extension of cognitive user simulators that adds these three mechanisms as inspectable Bayesian components.
We stated two V\&V properties at the framework level (face validity by trust calibration, structural validity by ablation-based attribution) and demonstrated on the HC3 corpus that each mechanism contributes a distinct behaviour: the trust component calibrates posterior beliefs ($0.162$ vs $0.356$ TCE), the effort budget constrains verification load ($34.5\%$ vs $70.2\%$), and the strategic policy reduces high-stakes regret ($0.122$ vs $0.171$).
The same architecture applies to clinical decision support, predictive maintenance, and consumer recommendation under AI-curated lists.
The released detection-difficulty index and diagnostic harness allow other user-simulation work to inherit the provenance-aware components without re-implementing them.

\bibliographystyle{ACM-Reference-Format}
\bibliography{references}

\end{document}